# Aggregation and Emergence in Agent-Based Models: A Markov Chain Approach

Sven Banisch* & Ricardo Lima** & Tanya Araújo***

(*) Mathematical Physics, Bielefeld University, Germany
*sven.banisch@UniVerseCity.De*
(**) Dream & Science Factory, Marseilles (France)
(***) ISEG - Technical University of Lisbon (TULisbon) and Research Unit on Complexity in Economics (UECE), Portugal
*tanya@iseg.utl.pt*

**Abstract**

We analyze the dynamics of agent–based models (ABMs) from a Markovian perspective and derive explicit statements about the possibility of linking a microscopic agent model to the dynamical processes of macroscopic observables that are useful for a precise understanding of the model dynamics. In this way the dynamics of collective variables may be studied, and a description of macro dynamics as emergent properties of micro dynamics, in particular during transient times, is possible.

## 1 Introduction

Our work is a contribution to interweaving two lines of research that have developed in almost separate ways: Markov chains and agent–based models (ABMs). The former represents the simplest form of a stochastic process while the latter puts a strong emphasis on heterogeneity and social interactions. The usefulness of the Markov chain formalism in the analysis of more sophisticated ABMs has been discussed by Izquierdo et al. (2009), who look at 10 well–known social simulation models by representing them as a time–homogeneous Markov chain. Among these models are the Schelling segregation model (Schelling (1971)), the Axelrod model of cultural dynamics (Axelrod (1997)) and the sugarscape model from Epstein and Axtell (1996). The main idea of Izquierdo et al. (2009) is to consider all possible configurations of the system as the state space of the Markov chain. Despite the fact that all the information of the dynamics on the ABM is encoded in a Markov chain, it is difficult to learn directly from this fact, due to the huge dimension of the configuration space and its corresponding Markov transition matrix. The work of Izquierdo and co-workers mainly relies on numerical computations to estimate the stochastic transition matrices of the models.

Consider an ABM defined by a set $\mathbf{N}$ of agents, each one characterized by individual attributes that are taken from a finite list of possibilities. We denote the set of possible attributes by $\mathbf{S}$ and we call the *configuration space* $\mathbf{\Sigma}$ the set of all possible combinations of attributes of the agents, i.e., $\mathbf{\Sigma} = \mathbf{S}^N$. This also incorporates models where agents move on a lattice (e.g., in the sugarscape model) because we can treat the sites as "agents" and use an attribute to encode whether a site is occupied or not. The updating process of the attributes of the agents at each time step typically consists of two parts. First, a random choice of a subset of agents is made according to some probability distribution $\omega$. Then the attributes of the agents are updated according to a rule, which depends on the subset of agents selected at this time. With this specification, ABMs can be represented by a so–called random map representation which may be taken as an equivalent definition of a Markov chain (Levin et al. (2009)). Hence, ABMs are Markov chains on $\mathbf{\Sigma}$ with a transition matrix $\hat{P}$. For a class of ABMs we can compute transition probabilities $\hat{P}(x,y)$ for any pair $x, y \in \mathbf{\Sigma}$ of agent configurations. We refer to the process $(\mathbf{\Sigma}, \hat{P})$ as micro chain.

When performing simulations of an ABM we are actually not interested in all the dynamical details but rather in the behavior of variables at the macroscopic level (such as average opinion, number of communities, etc.). The formulation of an ABM as a Markov chain $(\mathbf{\Sigma}, \hat{P})$ enables the development of a mathematical framework for linking the micro–description of an ABM to a macro–description of interest. Namely, from the Markov chain perspective, the transition from the micro to the macro level is a projection of the Markov chain with state space $\mathbf{\Sigma}$ onto a new state space $\mathbf{X}$ by means of a (projection) map $\Pi$ from $\mathbf{\Sigma}$ to $\mathbf{X}$. The meaning of the projection $\Pi$ is to lump sets of micro configurations in $\mathbf{\Sigma}$ according to the macro property of interest in such a way that, for each $X \in \mathbf{X}$, all the configurations of $\mathbf{\Sigma}$ in $\Pi^{-1}(X)$ share the same property.

The price to pay in passing from the micro to the macrodynamics in this sense (Kemeny and Snell (1976), Chazottes and Ugalde (2003)) is that the projected system is, in general, no longer a Markov chain: long memory (even infinite) may appear in the projected system. In particular, well known conditions for lumpability (Kemeny and Snell (1976)) make it possible to decide whether the macro model is still Markov. Conversely, this setting can also provide a suitable framework to understand how aggregation may lead to the emergence of long range memory effects.

## 2 The Model

We illustrate these ideas at the example of the Voter Model (VM) (see Refs. Kimura and Weiss (1964); Castellano et al. (2009)). In the VM, $\mathbf{S} = \{0, 1\}$ meaning that each agent is characterized by an attribute $x_i, i = 1, \ldots, N$ which takes a value among two possible alternatives. The set of all possible combinations of attributes of the agents is $\Sigma = \{0, 1\}^N$, that is, the set of all bit–strings of length $N$. At each time step in the iteration process, an agent $i$ is chosen at random along with one of its neighboring agents $j$. If the states $(x_i, x_j)$ are not equal already, agent $i$ adopts the state of $j$ (by setting $x_i = x_j$). At the microscopic level of all possible configurations of agents the VM corresponds therefore to an absorbing random walk on the $N$–dimensional hypercube. It is well known that the model has the two absorbing states $(1, \ldots, 1)$ and $(0, \ldots, 0)$. For a system of three agents this is shown in Fig.1.

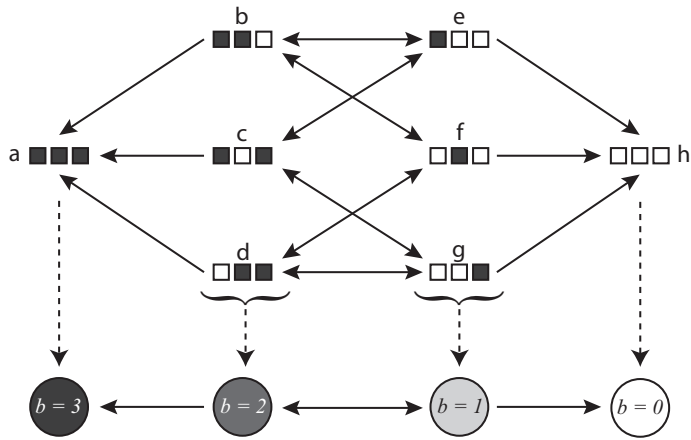

Figure 1: The micro chain for the VM with 3 agents and its projection onto a random walk obtained by agglomeration of states with the same number of black agents $b$.

Opinion models as the VM are a nice examples where our projection construction is particularly meaningful. There, we consider the projection $\Pi_b$ that maps each $x \in \Sigma$ into $X_b \in \mathbf{X}$ where $b$ is the number of agents in $x$ with opinion 1. The projected configuration space is then made of the $X_b$ where $0 \leq b \leq N$ (see Fig.1). Markov chain theory, in particular lumpability, allows us to determine conditions for which the macro chain on $\mathbf{X} = (X_0, \ldots, X_b, \ldots, X_N)$ is again a Markov chain. We find that this requires that the probability distribution $\omega$ must be invariant under the group $\mathcal{S}_N$ of all the permutations of $N$ agents and therefore uniform. This underlines the theoretical importance of homogeneous or complete mixing in the analysis of the VM and related models.

In this way our method enables the use of Markov chain instruments in the mathematical analysis of ABMs. In Markov chains with absorbing states (and therefore in the ABM) the asymptotic status is quite trivial. As a result, it is the understanding of the transient that becomes the interesting issue. In order to analyze the transient dynamics for the macro dynamics, all that is needed is to compute the fundamental matrix $\mathbf{F}$ of the Markov chain (Kemeny and Snell (1976)). For the binary VM we are able to derive a closed form expression for the elements in $\mathbf{F}$ for arbitrary $N$ which provides us with all the information needed to compute the mean quantities and variances of the transient dynamics of the model. In addition, we show in the VM with three opinion alternatives ($\mathbf{S} = \{0, 1, 2\}$) how restrictions in communica-tion (bounded confidence) lead to stable co–existence of different opinions because new absorbing states emerge in the macro chain (see Banisch et al. (2012)).

## 3 Some Results

The micro chains obtained via the random map representations help to understand the role of the collection of (deterministic) interaction rules used in the model from one side and of the probability distribution $\omega$ governing the sequential choice of the rules used to update the system at each time step from the other side. The importance of this probability distribution is to encode social relations and exchange actions. In our setting it becomes explicit how the symmetries in $\omega$ translate into symmetries of the micro chain. If we decide to remain at a Markovian level, then the partition, or equivalently the collective variables to be used to build the macro model must be compatible with the symmetry of the probability distribution $\omega$. In order to account for an increased level of heterogeneity the partition of the configuration space defining the macro–level has to be refined. A first result into this direction is that the symmetry group of agent permutations on $\omega$ informs us about ensembles of agent configurations that can be interchanged without affecting the probabilistic structure of micro chain. Consequently, these ensembles can be lumped into the same macro state and the dynamical process projected onto these states is still a Markov chain. It is clear, however, that, in absence of any symmetry, there is no other choice than to stay at the micro–level because no Markovian description at a macro–level is possible in this case.

In our opinion, a well posed mathematical basis for linking a micro–description of an ABM to a macro–description may help the understanding of many of the properties observed in ABMs and therefore provide information about the transition from the interaction of individual actors to the complex macroscopic behaviors observed in social systems. We summarize our main results below:

1. We formulate agent–based models as Markov chains at the micro level with explicit transition probabilities.

2. This allows the use of lumpability arguments to link between the micro and the macro level.

3. In case of a non–lumpable macro description this explains the emergence non–trivial dynamical effects (long memory).

4. In the Voter Model, homogeneous mixing leads to a macroscopic Markov chain which underlines the theoretical importance of homogeneous mixing.

5. This chain can be solved including mean convergence times and variances.

6. The stable co–existence of different opinions with in the bounded confidence model follows from the emergence of new absorbing states in the macro chain.

7. Heterogeneous mixing requires refinement and we show how to exploit the symmetries in the mixing distribution ($\omega$) to obtain a proper refinement.

# Further Reading

Banisch, S., Lima, R., and Araújo, T. (2012). Agent Based Models and Opinion Dynamics as Markov Chains. *accepted by Social Networks*, (http://arxiv.org/abs/1108.1716).

Banisch, S. and Lima, R. (2012). Markov Projections of the Voter Model. *unpublished*, (http://www.universecity.de/MarkovProjections0312.pdf).

# Acknowledgement

This work has benefited from financial support from the Fundação para a Ciência e a Tecnologia (FCT), under the *13 Multi-annual Funding Project of UECE, ISEG, Technical University of Lisbon*. Financial support of the German Federal Ministry of Education and Research (BMBF) through the project *Linguistic Networks* is also gratefully acknowledged (`http://project.linguistic-networks.net`).